\documentclass[aps,prd,preprint,superscriptaddress,tightenlines,nofootinbib]{revtex4}

\usepackage{graphicx}% Include figure files
\usepackage{dcolumn}% Align table columns on decimal point
\usepackage{bm}% bold math

\begin{document}

%\preprint line(s) will be ignored for PRL/PRD
%\preprint{CLEO Draft 05-05A} % For paper draft CBX YY-NN -> Draft YY-NNA
%\preprint{CLEO CONF YY-NN}   % For conference papers
%\preprint{ICHEP ABSnnn}      % For conference papers
\preprint{CLNS 05/1919}       % for CLNS notes
\preprint{CLEO 05-11}         % for CLNS notes
\title{Observation of the $h_c(^1P_1)$ State of Charmonium}
% for conference papers (ask CLEOAC for appropriate text)
%\thanks{Submitted to the 31$^{\rm st}$ International Conference on High Energy
%Physics, July 2002, Amsterdam}

%-------- INSERT HERE ------------
% Your author list goes here  REMOVE EVERYTHING to END INSERT and
% replace with your authorlist (ask cleoac).

\author{J.~L.~Rosner}
\affiliation{Enrico Fermi Institute, University of
Chicago, Chicago, Illinois 60637}
\author{N.~E.~Adam}
\author{J.~P.~Alexander}
\author{K.~Berkelman}
\author{D.~G.~Cassel}
\author{V.~Crede}
\author{J.~E.~Duboscq}
\author{K.~M.~Ecklund}
\author{R.~Ehrlich}
\author{L.~Fields}
% ...
%\author{(CLEO Collaboration)} %FOR PRD_SPECIAL_CHANGEME
\collaboration{CLEO Collaboration} %FOR PRL,CLNS
\noaffiliation

%-------- END INSERT ------------

%please hard code the date when you have a final draft and submit to CLEOAC
\date{May 23, 2005}

\begin{abstract} 

The $h_c(^1P_1)$ state of charmonium has been observed in the reaction 
$\psi(2S)\to\pi^0 h_c\to(\gamma\gamma)(\gamma\eta_c)$ using 3.08 million $\psi(2S)$ 
decays recorded in the CLEO detector.  Data have been analyzed both for the 
inclusive reaction, where the decay products of the $\eta_c$ are
not identified, and for exclusive reactions, in which $\eta_c$ decays are 
reconstructed in seven hadronic decay channels.  
We find $M(h_c)=3524.4 \pm 0.6 \pm
0.4$~MeV which corresponds to a hyperfine splitting $\Delta
M_{hf}(1P)\equiv \left<M(^3P_J)\right>-M(^1P_1)=+1.0\pm0.6\pm0.4\;\mathrm{MeV}$,
and 
$\mathcal{B}(\psi(2S)\to\pi^0
h_c)\times\mathcal{B}(h_c\to\gamma\eta_c)=(4.0 \pm 0.8 \pm 0.7) \times
10^{-4}$.
\end{abstract}
\maketitle

Over the past thirty years charmonium spectroscopy has provided valuable insight 
into the quark--antiquark interaction of quantum chromodynamics (QCD).  
QCD-based potential models have been quite successful in predicting masses, 
widths, and dominant decays of several charmonium states.  The central potential 
in most of these calculations is assumed to be composed of a vector Coulombic
potential ($\sim$$1/r$) 
and a scalar confining potential ($\sim$$r$). Under these assumptions, 
the spin-spin interaction in the lowest order is 
finite only for $L=0$ states.  It leads to the hyperfine splittings 
$\Delta M_{hf}(nS)\equiv M(n^3S_1)\!-\!M(n^1S_0)$ between spin-triplet 
and spin-singlet $S$-wave states of charmonium, which have been measured as
$\Delta M_{hf}(1S)=M(J/\psi)\!-\!M(\eta_c) = 115\!\pm\!2$~MeV~\cite{etacp},
$\Delta M_{hf}(2S)=M(\psi(2S))\!-\!M(\eta_c')=43\!\pm\!3$~MeV~\cite{etacp,etacp2}.  
It also leads to the prediction that the hyperfine 
splitting $\Delta M_{hf}(\left<M(^3P_J)\right>\!-\!M(^1P_1))$ for $P$-wave 
states should be zero.  Higher-order corrections are expected to provide no more 
than a few MeV deviation from this result~\cite{appel,godfreyrosner,dave}.  
Lattice QCD
calculations~\cite{lattice} predict $\Delta M_{hf}(1P)=+1.5$ to $+3.7$ MeV, but
with uncertainties at the few-MeV level.  Larger values of $\Delta M_{hf}(1P)$ could result if
the confinement potential had a vector component or if coupled channel effects
were important. In order to discriminate
between these possibilities, it is necessary to identify the $h_c(^1P_1)$ state
and to measure its mass to $\mathcal{O}(1\;{\rm MeV})$
as the mass of the $^3P_J$ centroid is very well known,
$\left<M(^3P_J)\right>=3525.36\!\pm\!0.06$~MeV~\cite{pdg},

In this Letter we report the successful identification of $h_c$ in the isospin-violating reaction
\begin{equation}
e^+e^- \to \psi(2S) \to \pi^0 h_c\;,\;h_c \to \gamma\eta_c\;,\;\pi^0 \to \gamma\gamma.
\label{eqn:casc}
\end{equation}
Two methods are used: one in which the $\eta_c$ decays are reconstructed
(exclusive), which has an advantage in signal purity, and the other
in which the $\eta_c$ is measured inclusively, which has larger signal
yield. Together these approaches provide a result of unambiguous
significance, 
and allow a precise determination of
the mass of $h_c$ and the branching fraction product
$\mathcal{B}_\psi \mathcal{B}_h$,
where $\mathcal{B}_\psi\equiv \mathcal{B}(\psi(2S)\to\pi^0 h_c)$ and $B_h\equiv 
\mathcal{B}(h_c\to\gamma\eta_c)$.
Theoretical estimates of the product
$\mathcal{B}_\psi \mathcal{B}_h$ vary by nearly two orders of
magnitude, $(0.5-40)\times10^{-4}$~\cite{godfreyrosner,dave}.

The Crystal Ball Collaboration at SLAC searched for $h_c$ using the reaction of Eq.~(\ref{eqn:casc}) but were only able to set a 95\% confidence 
upper limit
$\mathcal{B}_\psi \mathcal{B}_h<16\times10^{-4}$ in the 
mass range $M(h_c)=(3515-3535)$~MeV~\cite{cball}. The FNAL E760 Collaboration searched for $h_c$ in 
the reaction $p\bar{p}\to h_c\to\pi^0J/\psi,\;J/\psi\to e^+e^-$, and reported a statistically 
significant enhancement with $M(h_c)=3526.2\pm0.15\pm0.2$ MeV,
$\Gamma(h_c)\le1.1$ MeV~\cite{e760}.  The 
measurement was repeated twice by the successor experiment E835 with $\sim$2$\times$ and 
$\sim$3$\times$ larger luminosity, but no confirming signal for $h_c$ was observed in $h_c\to\pi^0 J/\psi$ decay~\cite{dave}.

A data sample consisting of $3.08\times10^6$ $\psi(2S)$ decays was obtained with the CLEO III and 
CLEO-c detector configurations~\cite{cleoii, dr, rich, yellowbook} 
at the Cornell Electron Storage Ring.  
The CLEO~III detector features a solid angle coverage for charged and
neutral particles of 93\%. The charged particle tracking system
achieves a momentum resolution of $\sim$0.6\% at 1 GeV, and the
calorimeter photon energy resolution is 2.2\% for $E_\gamma=1$ GeV and 5\% at 100 MeV.
Two particle identification systems, one based on energy-loss
($dE/dx$)
in the drift chamber and the other a ring imaging Cherenkov (RICH) detector, 
are used to distinguish pions from kaons. 

\begin{figure}[t]
\includegraphics*[width=\textwidth]{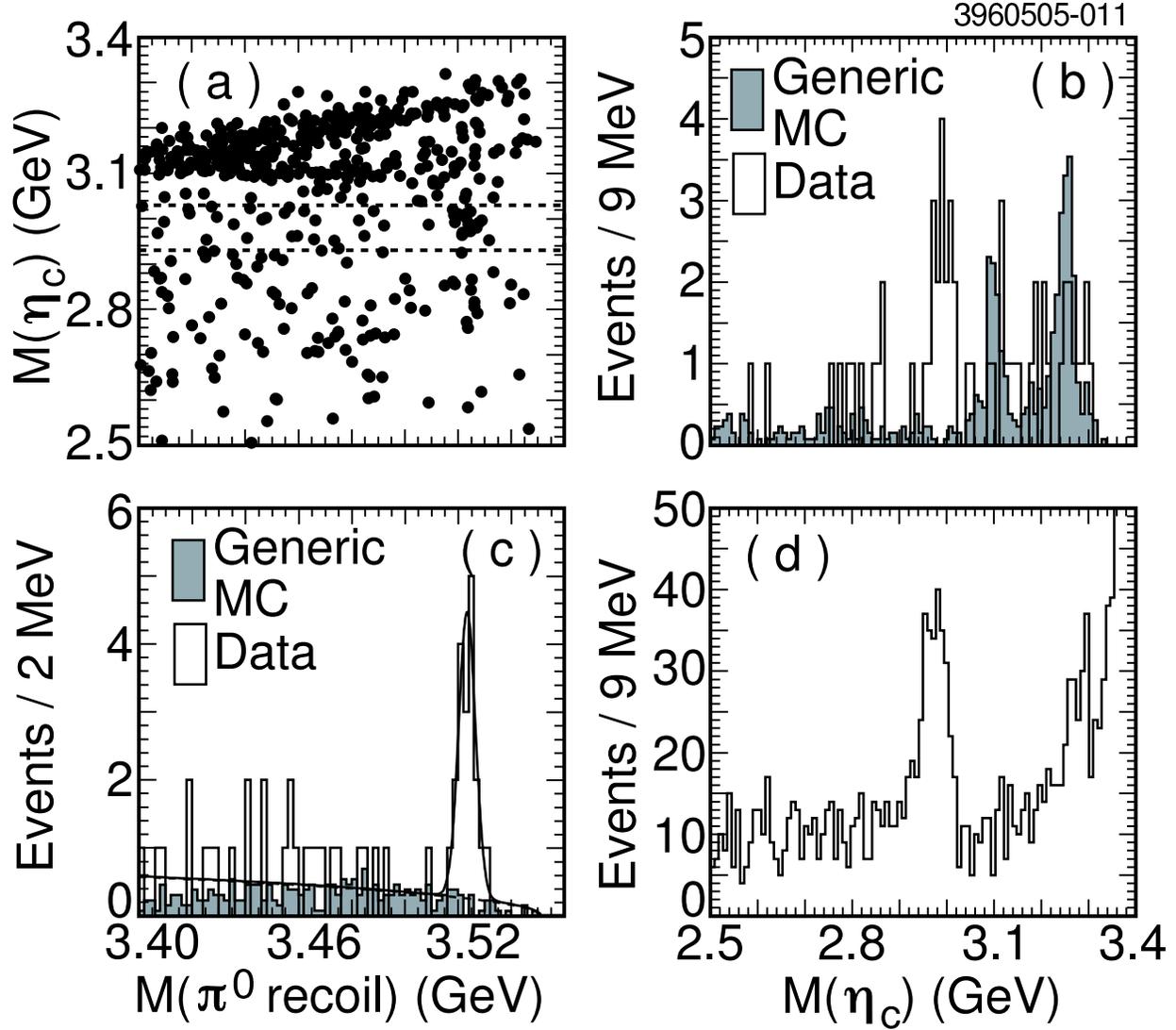}
\caption{
Exclusive analysis: 
(a) Scatter plot of the reconstructed
$\eta_c$ candidate mass vs.~the recoil mass against $\pi^{0}$ 
for data. The horizontal band in the vicinity of
$M(J/\psi)$ and the diagonal band at larger $\eta_c$ candidate mass
correspond to $\psi(2S)\rightarrow \pi^0\pi^0 J/\psi$ and
$\psi(2S)\rightarrow\gamma\chi_{c0}$, respectively.
The dashed lines denote the region $M(\eta_c)=2982\pm50$ MeV. 
Data events (open histograms) and Monte
Carlo background estimate (shaded histograms) of
(b) reconstructed $\eta_c$ candidate mass projection for
$M(\pi^0\;{\rm recoil})=3524\pm8$~MeV and 
(c) recoil $h_c$ candidate mass spectrum for $M(\eta_c)=2982\pm50$~MeV. The
peaks in b) near $M$=3.1~GeV and 3.25~GeV correspond to
$\psi(2S)\rightarrow \pi^0\pi^0 J/\psi$ and
$\psi(2S)\rightarrow\gamma\chi_{c0}$, respectively.
(d) Reconstructed $\eta_c$ candidate mass for data in the direct decay
$\psi(2S)\rightarrow\gamma\eta_c$. The peak near $M=3.4$~GeV is from the
direct decay $\psi(2S)\rightarrow\gamma\chi_{c0}$.}
\label{fig:excl}
\end{figure}

Half of the $\psi(2S)$ data were accumulated with a newer detector
configuration,
CLEO-c~\cite{yellowbook}, in which the silicon strip vertex detector was replaced
with an all-stereo six-layer wire chamber. The two detector
configurations also correspond to different accelerator lattices.
Studies of Monte Carlo simulations and the data reveal no
significant differences in the capabilities of the two detector configurations;
therefore the CLEO~III and
CLEO-c datasets are analyzed together.

The inclusive and exclusive analyses share a common initial sample of
events and numerous selection criteria.  Details of the
analyses are provided in a companion paper~\cite{hcprd}.
Event selection for both analyses require at least three electromagnetic showers 
and two charged tracks, each selected with standard CLEO criteria.  
For showers, $E_\gamma>30$ MeV is required.  
Candidates for $\gamma \gamma$ decays of $\pi^0$ or $\eta$
mesons satisfy the requirement that $M(\gamma\gamma)$
be within 3 standard deviations ($\sigma$) of the known $\pi^0$ or
$\eta$ mass, respectively.  These candidates are kinematically fit, 
constraining $M(\gamma\gamma)$ to the appropriate mass to improve
$\pi^0/\eta$ energy resolution. 
Charged tracks are required to have well-measured momenta 
and to satisfy criteria based on
the track fit quality. They must also be consistent with originating
from the interaction point in three dimensions.

\begin{figure}[t]
\includegraphics*[width=\textwidth]{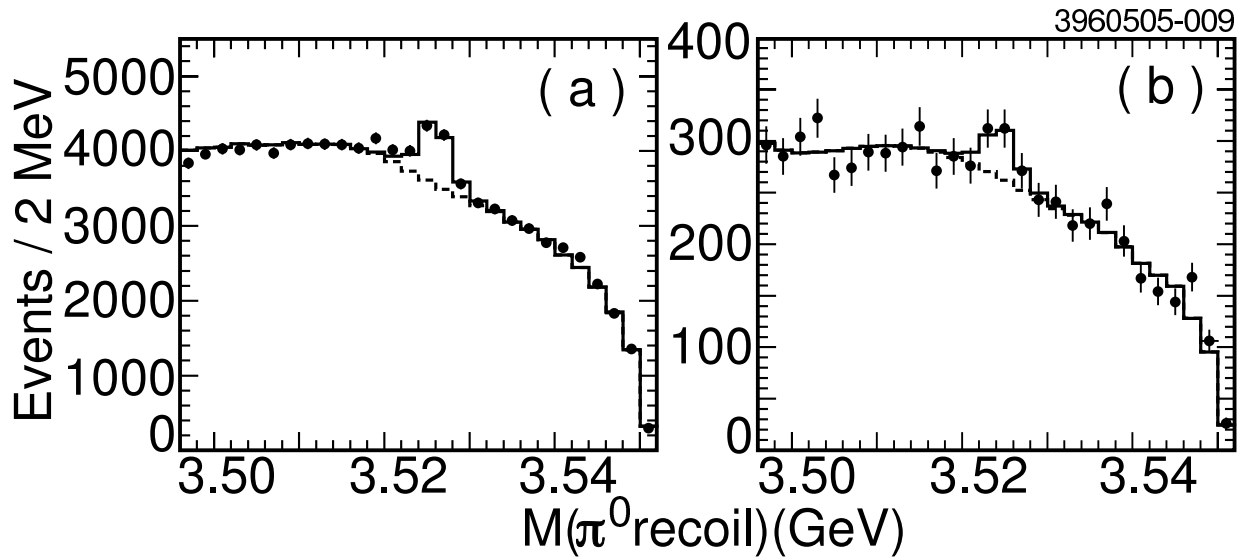}
\caption{
Inclusive analysis: Recoil mass against $\pi^0$ for 
(a) Monte Carlo sample for 39.1 million $\psi(2S)$ and (b) data for 3
million $\psi(2S)$.
See text for details.}
\label{fig:incl}
\end{figure}

Both techniques identify $h_c$ as an enhancement in the
spectrum of neutral pions 
from the reaction $\psi(2S)\to\pi^0h_c$~\cite{Aulchenko:2003qq}. 
For this purpose, it is
useful to remove neutral pions originating from any other
reaction. It is easy to remove most of the $\pi^0$ arising
from $\psi(2S)\to\pi^+\pi^-J/\psi$, with
$J/\psi\to \pi^0\;+$~hadrons and $\pi^0\pi^0J/\psi$, with $J/\psi\to$~any.
Since the recoil spectra against 
$M(\pi^+\pi^-)$ (both analyses) and $M(\pi^0\pi^0)$ (inclusive only) show prominent peaks for $J/\psi$; these events are removed by appropriate selection around $M(J/\psi)$.

In the exclusive analysis, $\eta_c$ are reconstructed
in seven channels: $K_S^0 K^\pm \pi^\mp$, $K_L^0 K^\pm \pi^\mp$, $K^+K^-\pi^+\pi^-$, 
$\pi^+\pi^-\pi^+\pi^-$, $K^+K^-\pi^0$, $\pi^+\pi^-\eta(\to\gamma\gamma)$, and $\pi^+\pi^-\eta(\to\pi^+\pi^-\pi^0)$.  
The sum of the branching fractions is $(9.7\pm2.7)\%$~\cite{pdg}.  
The decay chain in Eq.~(\ref{eqn:casc}) as well as these $\eta_c$ decays are identified from reconstructed charged 
particles, $\pi^0$ and $\eta$ mesons.
For $\eta$ decays to $\pi^+\pi^-\pi^0$, the
three-pion invariant mass is required to be within 20~MeV of the 
nominal $\eta$ mass. 
The $K_S^0$ candidates are selected from pairs of oppositely charged and
vertex-constrained tracks with invariant mass within 10 MeV, roughly
4$\sigma$, of the $K_S^0$ mass.
A kinematically constrained 4C fit is performed for
each event. A 1C fit
is performed for the $\eta_c \to K_L^0 K^\pm \pi^\mp$ decay because
the $K_L^0$ is not detected. It is required that $M(\eta_c)=2980\pm 50$~MeV. No explicit
selection of the energy of the photon from $h_c \to \gamma\eta_c$ is required.
The final selection is on the $\eta_c$ candidate
mass; however, to improve resolution, the $h_c$ mass is calculated from the 4-momentum of the $\psi(2S)$ and the 
$\pi^0$ instead of the invariant mass of its decay products.

In addition to $\psi(2S)\to \pi\pi J/\psi$ decays discussed above,
a fraction of $\psi(2S)$ decays proceed through $\psi(2S)\to 
\pi^0J/\psi$ and $\psi(2S)\to \gamma\chi_{c_J} \to \pi^0 X$.  To suppress the $\pi^0$ background, each signal photon
candidate is paired with all other
photons in that event.  If the invariant mass of any pair is within the $\pi^0$ mass requirement, the
event is removed. 

Fig.~\ref{fig:excl}(a) shows the scatter plot of the $\eta_c$ candidate
mass versus $\pi^0$ recoil mass (sum of all channels).  
Many events are seen in the vicinity of $M(J/\psi)$. In the 
mass band $M(\eta_c)=2980\pm50$~MeV an enhancement of events is
observed at larger $\pi^0$ recoil mass.  The projection of the events in
this band and the Monte Carlo background estimate is shown 
in Fig.~\ref{fig:excl}(c). A prominent peak is clearly visible over a very
small background. The projection of the events in the mass
band $M(\pi^0\;{\rm recoil})=3524\pm8$~MeV and the Monte Carlo
background estimate, shown in Fig.~\ref{fig:excl}(b), indicate that most of these
events arise from $\eta_c$ decay.
The $\pi^0$ recoil mass spectrum, in Fig.~\ref{fig:excl}(b), is fit using a double Gaussian shape determined from Monte Carlo simulation and an 
ARGUS function background~\cite{argus}.  The maximum likelihood fit
yields $17.5\pm4.5$ counts in the peak and $M(h_c)=3523.6\pm0.9$ MeV.

Several different methods have been utilized to estimate the
statistical significance $s$ of the signal~\cite{hcprd}, including 
the fit to the recoil mass spectrum just described,
Poisson fluctuations of MC-predicted backgrounds
inside the signal window, and a binomial statistics calculation
using the assumption that the events in the recoil mass distribution
are 
uniformly
distributed. Using the difference between the likelihood values of the
fit
with and without the signal contribution, we obtain $s=6.1\sigma$;
similar calculations with different $\eta_c$ mass ranges yield
$s=5.5-6.6\sigma$. 
The probability that Poisson fluctuations of the background,
estimated from the generic MC sample,  
completely account for the observed events in the signal region
is $1\times10^{-9}$ ($s=6.0\sigma$). 
The binomial probability that the number of data events in
Fig.~\ref{fig:excl}(b) and \ref{fig:excl}(c)
fluctuate to be
greater than the number of events in the signal region
is $2.2\times
10^{-7}$, corresponding to $s\sim$5.2$\sigma$.

To test our ability to reconstruct $\eta_c$ decays and provide normalization for the branching
fraction measurement, $\mathcal{B}_\psi \mathcal{B}_h$, 
the direct radiative decay $\psi(2S)\to\gamma\eta_c$ is studied.  Events 
are reconstructed in the same $\eta_c$ decay channels as for the
$h_c$ search,
but with much better yields.
Relative yields among the various channels are similar to previous results~\cite{pdg} and the
$\eta_c$ peak shape was verified for each channel.
Figures~\ref{fig:excl}(b) and \ref{fig:excl}(d) show the reconstructed
mass spectra for the $\eta_c$ candidates from $h_c$ and direct $\psi(2S)$
decay, respectively.
The $\eta_c$ mass resolution in the photon recoil mass spectrum is identical for
all seven channels. This distribution summed over all channels (not shown)
is fit using a peak shape which consists of a Monte 
Carlo-derived double Gaussian 
convolved with a Breit-Wigner function (with $M(\eta_c)=2979.7$ MeV, $\Gamma(\eta_c)=27$ MeV).  
It yields $220\pm22$ counts.
The efficiency-corrected ratio of $h_c$ decays to direct decays, which
corresponds to
$\mathcal{B}_\psi \mathcal{B}_h/\mathcal{B_D}$, 
where $\mathcal{B_D}\equiv \mathcal{B}(\psi(2S)\to\gamma \eta_c)$, is determined to be
$0.178\pm0.049$. The CLEO~\cite{etacbr} and PDG~\cite{pdg} values are
combined to obtain $\mathcal{B_D} = (0.296\pm0.046)\%$.
Multiplying these two results yields
$\mathcal{B}_\psi \mathcal{B}_h=(5.3\pm1.5)\!\times\!10^{-4}$ from the exclusive analysis.

\begin{figure}[t]
\includegraphics*[width=\textwidth]{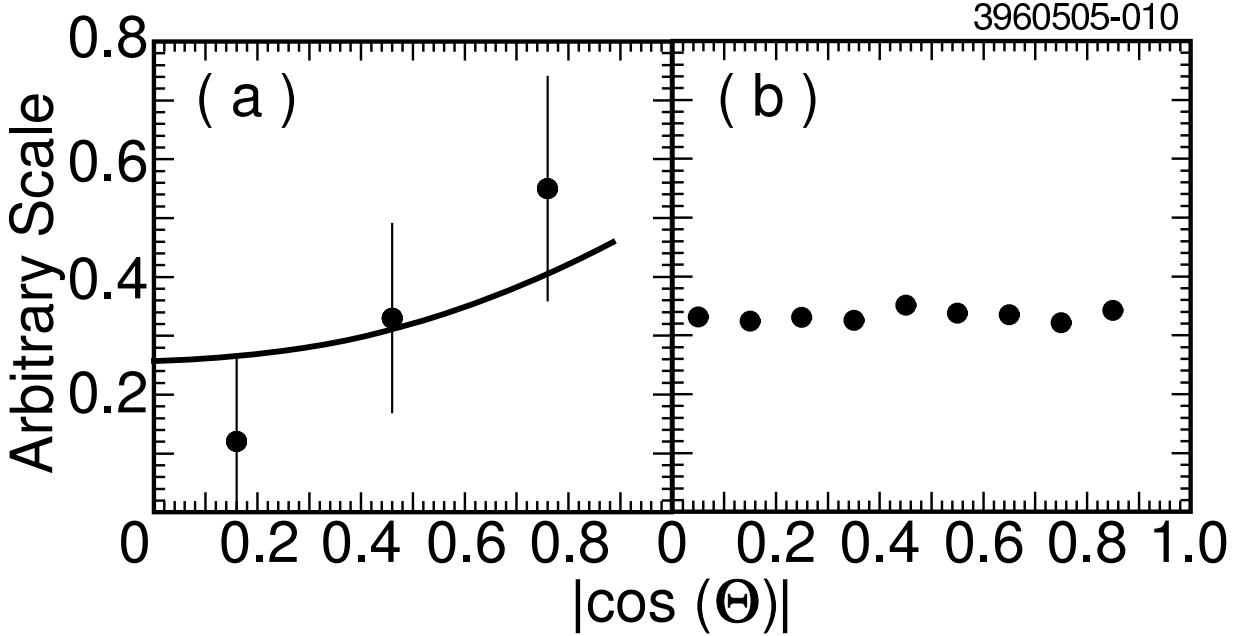}
\caption{
Inclusive analysis: Efficiency-corrected fit yields versus
$|\cos\theta|$ for data with $E_\gamma=503\pm35$~MeV, (a) for $h_c$
yield, the curve corresponds to the best fit
$\propto(1+\cos^2\theta)$ and
(b) for the nearly isotropic background yield.}
\label{fig:cos}
\end{figure}

In the inclusive analysis, we explore two methods
to enhance the selection of neutral pions which are part of the chain 
$\psi(2S)\to\pi^0h_c\to\pi^0\gamma\eta_c$.  One way is to specify that there be only one 
photon in the event with energy for the transition $h_c\to\gamma\eta_c$, $E_\gamma\approx503$ MeV
(corresponding to $M(h_c)\approx3526$ MeV).  
Another way is to specify 
that the mass recoiling against the photon and $\pi^0$
for the event should be near the mass of $\eta_c$. Both approaches are investigated, leading to 
consistent results, as detailed in Ref.~\cite{hcprd}.

A combined sample of generic $\psi(2S)$ decay and signal Monte Carlo events is used to optimize the 
criteria for the final event selection.
The resulting selection criteria determined were $E_\gamma=503\pm35$ MeV for hard photon acceptance 
in one approach and $M(\eta_c)=2980\pm35$ MeV in the other.
As a result of the Monte Carlo studies, a number of selection criteria, in which the two approaches occasionally differ, are made.  
These include requiring only one $\pi^0$ in the signal region,
removing hard photons that reconstruct $\eta$ mesons 
with any other photon, accepting photons in the calorimeter endcaps, removing photons from the 
cascade reaction $\psi(2S)\to\gamma\chi_J\to\gamma\gamma J/\psi$, and the choice of the background shape.

\begin{table}[t]
\caption{\label{tab:results}
Results for the inclusive and exclusive analyses for the
reaction $\psi(2S)\to\pi^0h_c\to\pi^0\gamma\eta_c$.  First errors are
statistical, and the second errors are systematic, as described in the
text and Table~\ref{tab:systematics}.}
\begin{center}
\setlength{\tabcolsep}{8pt}
\begin{tabular}{lcc}
\hline \hline
 & Inclusive & Exclusive \\
\hline
Counts & $150\!\pm\!40$ & $17.5\!\pm\!4.5$ \\
Significance & $\sim$3.8$\sigma$  & $6.1\sigma$ \\
$M(h_c)$ (MeV) & $3524.9\!\pm\!0.7\!\pm\!0.4$  & $3523.6\!\pm\!0.9\!\pm\!0.5$   \\
$\mathcal{B}_\psi \mathcal{B}_h$ $(10^{-4})$& $3.5\!\pm\!1.0\!\pm\!0.7$ & $5.3\!\pm\!1.5\!\pm\!1.0$ \\
\hline \hline
\end{tabular}
\end{center}
\end{table}

The recoil spectrum for the total Monte Carlo sample of 39.1 million
$\psi(2S)$ 
(13 times the size of the data sample), 
obtained in the $E_\gamma$-selection
approach with its optimized selection criteria,
is shown in Fig.~\ref{fig:incl}(a).  A product branching fraction 
$\mathcal{B}_\psi \mathcal{B}_h=4\times 10^{-4}$ was assumed.
The corresponding plot from the other approach is very similar.  
The $h_c$ signal is evident.  The overall efficiencies determined from the 
Monte Carlo sample are $13.4\%$ and $14.6\%$ for the two inclusive approaches.  Input values 
of $M(h_c)$ and $\mathcal{B}_\psi \mathcal{B}_h$ are well reproduced.
Results of Monte Carlo studies lead to the conclusion that the resonance fits to the data may be expected to have significance levels of 
$\sim$$4\sigma$, statistical error on the mass of $\sim$$\pm0.6$ MeV, 
and central values of the mass are reproduced within $\sim$$\pm0.6$ MeV
of the generated $M(h_c)$.

Figure~\ref{fig:incl}(b) shows the data and the fit using the Monte Carlo optimized 
criteria for the same inclusive approach as in Fig~\ref{fig:incl}(a).  
Features in the Monte Carlo such as signal width, signal to background ratio, and 
approximate background shape mirror the data faithfully.  The recoil 
spectrum and the fit for the other inclusive approach are very similar.  
Fit significance is approximately 3.8$\sigma$.
Results from the two inclusive approaches differ by small amounts, 
with differences from the averages in $M(h_c)$ of $\pm0.5$ MeV and in 
$\mathcal{B}_\psi \mathcal{B}_h$ of $\pm0.05\!\times\!10^{-4}$. The average
results are listed in Table~\ref{tab:results}.

The $h_c$ yield from the recoil mass against $\pi^0$ in the inclusive
analysis is studied as a function of the angular distribution of the
$h_c\to\gamma\eta_c$ photon. The $h_c$ yield, shown in
Fig.~\ref{fig:cos}(a), is found to follow a $1+\cos^2\theta$ distribution 
($\chi^2/d.o.f. = 1.7/2$) as expected for an E1 transition from a spin 1 state. 
The background yield, shown in Fig.~\ref{fig:cos}(b), is uniform in $\cos\theta$.
The $h_c$ yield in the exclusive analysis is not sufficient to draw
any conclusions regarding the corresponding angular distribution.

Systematic uncertainties 
in the two analyses due to various possible sources have been 
estimated.  Many sources are common, such as choice of background parameterization, 
$h_c$ resonance intrinsic width ($\Gamma=0.5-1.5$ MeV), $\pi^0$ line shape, 
bin size, and fitting range.  
The uncertainty in the branching ratio for $\psi(2S) \to \gamma
\eta_c$
enters the systematic uncertainty for the exclusive analysis only
while
the uncertainty on the number of $\psi(2S)$ decays applies to the
inclusive analysis only.
The estimated contributions are listed in Table~\ref{tab:systematics}.  
For the inclusive (exclusive) analysis they sum in quadrature to $\pm0.4$ (0.5) MeV in $M(h_c)$ and 
$\pm0.7\,(1.0)\times10^{-4}$ in
$\mathcal{B}_\psi \mathcal{B}_h$.
The largest systematic error for the exclusive analysis, 
$\mathcal{B}(\psi(2S)\to\gamma\eta_c)$, cancels in the ratio and we obtain
$\mathcal{B}_\psi \mathcal{B}_h/\mathcal{B_D}=0.178\pm0.049\pm0.018$.

\begin{table}[t]
\caption{\label{tab:systematics}
Summary of estimated systematic uncertainties and their sums in quadrature.  
N/A means not applicable.}
\begin{center}
\begin{tabular}{lcccc}
\hline \hline
 &  \multicolumn{2}{c}{$M(h_c)$ (MeV)} & \multicolumn{2}{c}{$\mathcal{B}_\psi  \mathcal{B}_h\times10^4$}\\
Systematic Uncertainty & ~~Incl.~~ & ~~Excl.~~ & ~~Incl.~~ & ~~Excl.~~ \\
\hline
Number of $\psi(2S)$ & N/A & N/A & 0.1 & N/A\\
$\mathcal{B}(\psi(2S)\to\gamma\eta_c)$ & N/A & N/A & N/A & 0.8\\ \hline
Background shape & 0.3 & 0.2 & 0.2 & 0.3\\
$\pi^0$ energy scale  & \multicolumn{2}{c}{0.2} & $\sim$0 & 0.1 \\
Signal shape & 0.1 & 0.1 & 0.3 & 0.2\\
$h_c$ width & 0.1 & 0.1 & 0.3 & 0.2\\
$\pi^0$ efficiency  & $\sim$0 & $\sim$0 & 0.1  &  0.3 \\
photon efficiency  & $\sim$0 & $\sim$0 & 0.2 & 0.2 \\
Binning, fitting range & 0.1 & 0.1 & 0.3 & 0.2\\
Modeling of $h_c$ decays & 0.1 &  0.3 & 0.3 & $\sim$0\\
$\eta_c$ mass & 0.1 & 0.2 & 0.1 & 0.1 \\
$\eta_c$ width & $\sim$0 & $\sim$0 & 0.2 & 0.1 \\
$\eta_c$ branching ratios & N/A & $\sim 0$ & N/A & 0.1 \\
\hline
Sum in quadrature & $\pm$0.4 & $\pm$0.5 & $\pm$0.7 & $\pm$1.0\\
\hline \hline
\end{tabular}
\end{center}
\end{table}

To summarize, we have observed the $h_c$ state, the $^1P_1$ state of charmonium,
in the reaction $\psi(2S)\to\pi^0h_c,\;h_c\to\gamma\eta_c$, in exclusive 
and inclusive analyses. The significance of our observation is greater than 
$5\sigma$ under a variety of methods to evaluate this quantity. We
combine the results of the exclusive and inclusive analyses to
obtain $M(h_c)=3524.4 \pm 0.6 \pm 0.4$~MeV and $\mathcal{B}(\psi(2S)\to\pi^0
h_c)\times\mathcal{B}(h_c\to\gamma\eta_c)=(4.0 \pm 0.8 \pm 0.7) \times 10^{-4}$.
The following value is obtained for the hyperfine splitting:
$$\Delta M_{hf}(\left<M(^3P_J)\right>\!-\!M(^1P_1))=+1.0\pm0.6\pm0.4\;\mathrm{MeV.}$$
Thus, the combined result for $M(h_c)$ is consistent with the
spin-weighted average of the $\chi_{cJ}$ states. 

We gratefully acknowledge the effort of the CESR staff 
in providing us with excellent luminosity and running conditions.
This work was supported by the National Science Foundation
and the U.S. Department of Energy.
\newline\newline

\end{document}